\documentclass[runningheads]{llncs}
\usepackage[T1]{fontenc}
\usepackage{graphicx}
\usepackage{hyperref}

\begin{document}
%
\title{A Multi-Agent System for Semantic Mapping of Relational Data to Knowledge Graphs}
%
%
\author{Milena Trajanoska \and Riste Stojanov \and
Dimitar Trajanov}
\authorrunning{M. Trajanoska, R. Stojanov, and D. Trajanov}
%
\institute{University Ss. Cyril and Methodius, Skopje, N.Macedonia
\email{\{milena.trajanoska,riste.stojanov,dimitar.trajanov\}@finki.ukim.mk}}
\maketitle              
\begin{abstract}
    Enterprises often maintain multiple databases for storing critical business data in siloed systems, resulting in inefficiencies and challenges with data interoperability. A key to overcoming these challenges lies in integrating disparate data sources, enabling businesses to unlock the full potential of their data. Our work presents a novel approach for integrating multiple databases using knowledge graphs, focusing on the application of large language models as semantic agents for mapping and connecting structured data across systems by leveraging existing vocabularies. The proposed methodology introduces a semantic layer above tables in relational databases, utilizing a system comprising multiple LLM agents that map tables and columns to Schema.org terms. Our approach achieves a mapping accuracy of over 90\% in multiple domains.

\end{abstract}
\section{Introduction}

In today's data-driven enterprises, information is often distributed across multiple siloed databases, limiting the ability to fully leverage its value. Achieving seamless integration between heterogeneous data sources remains a major challenge, particularly due to differences in schema design, semantics, and data representation. Knowledge Graphs (KGs) \cite{hogan2021knowledge} have emerged as a powerful paradigm for bridging these gaps by creating a unified, semantic view over disparate datasets.

Traditional data integration efforts, however, often require extensive manual work to align schemas and define mappings. Recent advances in Large Language Models (LLMs) \cite{chang2024survey} provide new opportunities to automate complex reasoning tasks, including semantic mapping and knowledge graph construction. LLMs have shown impressive performance in tasks that involve understanding relationships between concepts, making them ideal candidates for assisting in structured data integration.

In this paper, we present a novel semantic multi-agent system that leverages LLM agents to automate the integration of relational databases into a unified knowledge graph. Our approach introduces a semantic layer above relational data by mapping tables and columns to concepts from the Schema.org vocabulary \cite{schemaorg}. Each agent in the system has a dedicated role: mapping relational structures to semantic concepts, establishing relationships between entities, and validating the generated knowledge graph.

We evaluate our system using the Spider dataset \cite{yu2018spider}, a complex benchmark featuring databases from diverse domains. By mapping real-world relational schemas to Schema.org concepts, we demonstrate that our multi-agent system effectively automates the semantic integration process, achieving high mapping accuracy across several application domains. Our results highlight the potential of LLM-driven semantic agents to significantly reduce the manual burden of data integration and to build more interoperable enterprise knowledge bases.

\section{Literature review}
Recent research has explored various approaches to knowledge graph (KG) and ontology construction, and data integration, particularly in enterprise and industry contexts.

The Virtual Knowledge Graph (VKG) paradigm \cite{xiao2019virtual} offers an alternative to traditional data integration approaches by providing flexible, virtual graphs that embed domain knowledge over existing data sources. The approach provides tools for Ontology-based data access, which have significant use-cases in multiple industry applications. 

In the context of Industry 4.0 (I4.0) where different systems exist, the demand for the creation of an integrated view increases. One such initiative is the Bosch Industry 4.0 Knowledge Graph (BI40KG) \cite{grangel2020knowledge} which was developed with the goal of integrating data from different sources, improving interoperability and traceability across manufacturing systems. 

Many efforts have been made for developing enterprise knowledge graphs (EKGs), in line with this, an Enterprise Knowledge Graphs Framework \cite{galkin2016integration} was created, which aims to bridge the gap between the increasing need for EKGs and the lack of formal methods for realising them. 

The use of Large Language Models (LLMs) in structured information extraction has gained significant traction recently. Findings suggesting that LLMs excel at reasoning tasks, and are being applied to information extraction tasks more often \cite{xu2024large}. 

Our previous work in the field compares general-purpose LLMs like ChatGPT with specialized models such as REBEL for joint entity and relation extraction \cite{trajanoska2023enhancing}. Using sustainability-related texts, pipelines for automatic Knowledge Graph construction were built. Results show that LLMs can improve accuracy in extracting structured knowledge, and foundational models also show promise for automated ontology generation.

\section{Methodology}
This section describes the methodology for creating the semantic multi-agent system as well as the data source used in evaluating the approach. The section is divided into the following subsections: Data description - stating the dataset used for evaluation and the semantic vocabularies, Graph-vector store constuction -  detailing the process of constructing a hybrid vector store containing subgraphs in a vector representation, and Multi-agent system architecture - detailing the entire flow of the mapping system along with the AI agents' roles.

\subsection{Data description}

The Yale Spider database\cite{yu2018spider} is a benchmark dataset used in semantic parsing and text-to-SQL tasks, featuring real-world relational data from various domains. The Spider challenge aims to create natural language interfaces for databases spanning multiple domains. It includes 10,181 questions and 5,693 distinct complex SQL queries across 200 databases with multiple tables, representing 138 domains. 

In this study, the database serves as the foundation for evaluating our multi-agent approach, allowing us to demonstrate the effectiveness of mapping relational data to a knowledge graph.

For creating the mappings, we utilize existing semantic vocabularies, which are defined and maintained by established organizations over a long time period. The reason for this is to avoid hallucinating terms that do not have definitions in any well-known ontologies. 

With the goal of including as many general concepts as possible, we use the Schema.org vocabulary \cite{schemaorg}, which represents a collection of schemas (types and properties) that can be used to markup content on the Web, enabling search engines to better understand the context of Web pages. It describes entities, their relationships, and actions, and can be readily expanded using a clearly defined extension framework. As of 2024, more than 45 million websites have used Schema.org to annotate their pages, resulting in over 450 billion structured data objects \cite{schemaorg}.

\subsection{Graph-vectore store construction}
 To enable our use case for generating correct semantic mappings, we have created a custom graph-vector index for retrieval augmented generation \cite{lewis2020retrieval} using the term definitions in Schema.org. 
 
\begin{figure}[!h]
        \centering
        \includegraphics[width=0.7\linewidth]{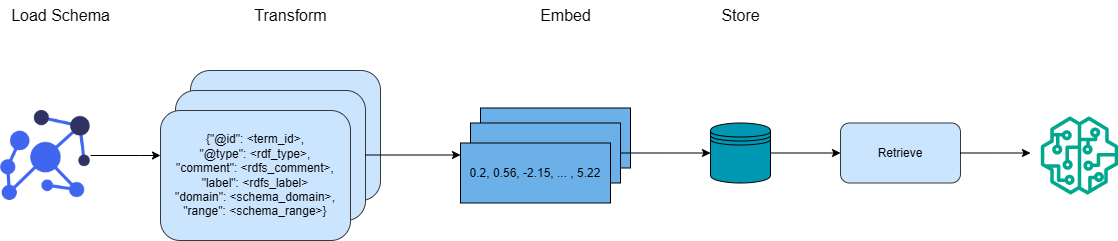}
        \caption{Graph - Vector store construction with ontology sub-graphs}
        \label{fig:vector_store}
    \end{figure}

 The index creation process is depicted in Figure \ref{fig:vector_store}. For each term, we extract its URI, type, human-readable comment, and label, as well as the domain and range for property terms. 
 Based on this information, we construct a one-hop subgraph for each term. This enriched subgraph offers a more detailed semantic context. Finally, each subgraph is transformed into a vector representation using a configurable text embedding transformer model \cite{patil2023survey}.

 The generated vectors are stored in a vector store. The subgraphs are retrieved from the vector store during the table mapping process, depending on their semantic similarity to the columns and values in the table being mapped.

\subsection{Multi-agent system architecture}
 The system consists of multiple GPT-4o-mini \cite{hurst2024gpt} agents with different responsibilities:  
    \begin{enumerate}
        \item \textbf{Mapping agent}: Maps columns to corresponding Schema.org terms.
        \item \textbf{Relation agent}: Determines primary and foreign keys in tables, creating links between entities in the knowledge graph.
        \item \textbf{Validator agent}: Validates and corrects the mapping and relations produced by the two previous agents.
    \end{enumerate}

\begin{figure}[!h]
        \centering
        \includegraphics[width=0.7\linewidth]{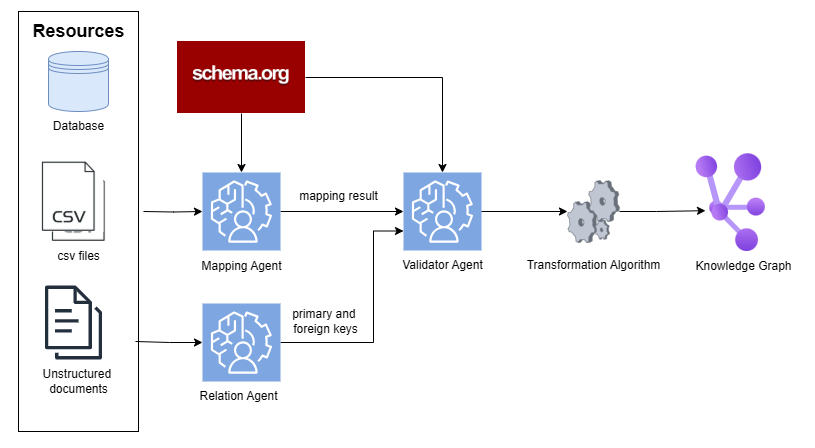}
        \caption{Data flow of the semantic multi-agent system}
        \label{fig:semantic_agents}
    \end{figure}

The mapping process is fully represented in Figure \ref{fig:semantic_agents}. The process operateos in a table-wise fashion. For each table, the Mapping agent receives the table name, all its column names, summary statistics for each column's values, a sample of the first K rows, and any available textual descriptions for the table or its columns. 

In addition, the mapping LLM is provided with the most similar terms from Schema.org to the table and column in order to map them accordingly. The most similar terms are retrieved from the custom graph-vector store described in the previous subsection.

In the subsequent step, all mapped tables are integrated into a unified knowledge base by accurately identifying primary and foreign key relationships through the Relation agent. This structured integration enables seamless interaction with the knowledge base by both internal and external systems.

In the final step, the Validator agent inspects and refines each table mapping and the final relation mapping to ultimately produce a more accurate result. The Validator agent may remove or re-map any relations it deems necessary to improve the accuracy. 

All the prompts for the different types of agents are available in our GitHub repository \cite{milena_trajanoska_knowledge_graph_agents} under the prompts folder.

All of the agents in the system output a confidence variable which can be one of the three following cateogories: HIGH, MEDIUM or LOW, to describe their confidence in the completion of their dedicated task. The categories are averaged from all steps to produce the final confidence class.

\section{Results and discussion}
The mapping system was evaluated on multiple databases from different domains present in the Spider data source. The accuracy of the generated mappings was assessed through manual validation, evaluating whether the Schema.org terms chosen by the system correctly reflected the semantics of the corresponding database elements. The evaluation dataset can be found in our GitHub repository \cite{milena_trajanoska_knowledge_graph_agents_eval} in the eval folder.

This section summarizes the findings for the entire execution of the system, including the time to generate the mapping result, as well as the accuracy of the mapping process.

Table \ref{tab:res_exec} shows that the execution time generally increases with the number of tables and columns being mapped. This is likely due to the complexity of larger tables and databases in the relationships they model.

\begin{table}[h!]
\caption{System execution time based on the number of tables and columns being mapped.}\label{tab:res_exec}
\begin{tabular}{|r|r|r|}
\hline
\multicolumn{1}{|l|}{\textbf{Number of tables}} & \textbf{Number of columns} & \multicolumn{1}{l|}{\textbf{Execution time (sec)}} \\ \hline
3                                               & 10                         & 122                                                \\ \hline
6                                               & 25                         & 179                                                \\ \hline
6                                               & 31                         & 191                                                \\ \hline
11                                              & 42                         & 342                                                \\ \hline
12                                              & 47                         & 295                                                \\ \hline
\end{tabular}
\end{table}

Concerning the validity of the generated mappings, we have conducted an evaluation for five different fields, namely: retail, movies, automotive, apartments, and delivery. We gathered all the datasets from these domains present in the Spider database and manually evaluated the accuracy of the mappings in each domain. 

The accuracy of the multi-agent system across the five domains is reported in Table \ref{tab:res_acc}. Apart from the mapping result, the multi-agent system returns the confidence of the mapping process, which can be one of the following categories: HIGH, MEDIUM, and LOW. For all confidence levels, the results are displayed in the table. 

\begin{table}[!h]
\caption{Accuracy of the multi-agent system on the five chosen domains presented by human evaluators}\label{tab:res_acc}
\begin{tabular}{|l|r|r|r|r|}
\hline
\multicolumn{1}{|c|}{\textbf{Sector}} & \multicolumn{1}{c|}{\textbf{Overall (\%)}} & \multicolumn{1}{l|}{\textbf{HIGH (\%)}} & \multicolumn{1}{l|}{\textbf{MEDIUM (\%)}} & \multicolumn{1}{l|}{\textbf{LOW (\%)}} \\ \hline
Retail                                & 78.72                                               & 95.83                                              & 64.29                                                & 55.56                                             \\ \hline
Movies                                & 90                                                  & 85.71                                              & 100.00                                               & /                                                 \\ \hline
Appartments                           & 93.54                                               & 100.00                                             & 88.89                                                & 50.00                                             \\ \hline
Automotive                            & 84                                                  & 100.00                                             & 80.00                                                & 25.00                                             \\ \hline
Delivery                              & 80.95                                               & 96.30                                              & 66.67                                                & 0                                                 \\ \hline
\end{tabular}
\end{table}

The highest accuracy (93.54\%) was observed in the 'Apartments' domain, while the lowest (78.72\%) occurred in the 'Retail' domain. Even for retail, the score is still satisfactory. It is evident that the mappings with certainty equal to HIGH are, in general, the most correct ones.

\section{Conclusion}

This paper introduced a semantic multi-agent system for enterprise data integration using Knowledge Graphs and LLMs. By aligning relational schemas with Schema.org, our approach enhances interoperability and enables unified, semantically rich graphs across domains.

LLM agents were assigned specialized roles—mapping, extraction, and validation—resulting in a scalable and automated integration pipeline. Experiments on the Spider benchmark showed high mapping accuracy and cross-domain applicability.

The main contributions of this work are:
\begin{itemize}
    \item The design and implementation of a \textbf{semantic multi-agent system} for automating relational-to-graph data integration.
    \item The innovative use of \textbf{LLM agents} to perform semantic mapping and reasoning over structured data using existing vocabularies.
    \item The empirical evaluation on a real-world benchmark, validating the system’s practicality and generalization capability.
\end{itemize}

Future work includes extending the system to support custom domain ontologies, enhancing the reasoning capabilities of the agents through fine-tuning, and investigating scaling strategies for integrating larger and more heterogeneous datasets.

\section{Acknowledgements}
This publication is based upon work from COST Action CA23147 GOBLIN - Global Network on Large-Scale, Cross-domain and Multilingual Open Knowledge Graphs, supported by COST (European Cooperation in Science and Technology, \href{https://www.cost.eu}{https://www.cost.eu}.

\bibliographystyle{splncs04}
\bibliography{bibliography}

\end{document}